\documentclass[journal]{IEEEtran}
\usepackage{graphicx}

\begin{document}
\thispagestyle{empty}
%

\title{Front-End Board with Cyclone\textsuperscript{\textregistered}  V 
as a Test High-Resolution Platform for the Auger\_Beyond\_2015 Front End Electronics}
\author{Zbigniew~Szadkowski,~\IEEEmembership{Member,~IEEE} for the Pierre Auger Collaboration
\thanks{Manuscript received June 07, 2014.  

This work was supported by the Polish National Center for
Research and Development under NCBiR Grant No. ERA/NET/ASPERA/02/11
and by the National Science Centre (Poland) under NCN Grant No. 2013/08/M/ST9/00322}
\thanks{Zbigniew~Szadkowski is with the University of \L{}\'od\'{z}, Department of Physics and Applied Informatics, 
Faculty of High-Energy Astrophysics, 90-236 \L{}\'od\'{z}, Poland,
(e-mail: zszadkow @kfd2.phys.uni.lodz.pl, phone: +48 42 635 56 59).}%
}

\maketitle

\begin{abstract}

The surface detector (SD) array of the Pierre Auger Observatory containing at present 1680 water Cherenkov detectors spread
over an area of 3000 km$^2$ started to operate since 2004. The currently
used Front-End Boards are equipped with no-more produced ACEX\textsuperscript{\textregistered} 
and obsolete Cyclone\textsuperscript{\textregistered}
FPGA (40 MSps/15-bit of dynamic range). 

Huge progress in electronics and new challenges from physics impose a significant
upgrade of the SD electronics either to improve a quality of measurements 
(much higher sampling and much wider dynamic range) or pick-up from a background extremely rare events
(new FPGA algorithms based on sophisticated approaches like e.g. spectral triggers or neural networks). 
Much higher SD sensitivity is necessary to confirm or reject
hypothesis critical for a modern astrophysics.

The paper presents the Front-End Board (FEB) with the biggest
Cyclone\textsuperscript{\textregistered}  V E FPGA 5CEFA9F31I7N,
supporting 8 channels sampled with max. 250 MSps @ 14-bit resolution.
Considered sampling for the SD is 120 MSps, however, the FEB has been developed
with external anti-aliasing filters to keep a maximal flexibility. Six channels are targeted to the SD,
two the rest for other experiments like: Auger Engineering Radio Array and additional muon counters.

The FEB is an intermediate design pluged-in the actually used Unified Board communicating with 
micro-controller at 40 MHz, however providing even 250 MSPs sampling with 20-bit dynamic range,
equipped in  a virtual NIOS\textsuperscript{\textregistered} processor and supporting
256 MB of SDRAM as well as with an implemented spectral trigger based on the Discrete Cosine Transform
for a detection of very inclined "old" showers. The FEB can also support a neural network developing
for a detection of "young" showers, potentially generated by neutrinos. 
\end{abstract}


%
\IEEEpeerreviewmaketitle

\section{Introduction}

\IEEEPARstart{T}{he} data taken with the Pierre Auger Observatory \cite{PAO}-
\cite{modulation} have lead to a number of major
breakthroughs in the field of ultra-high energy cosmic rays e.g. : a suppression of
the cosmic ray flux at energies above $5.5\cdot10^{19}$ eV has been doubtlessly established, 
top-down source processes such as the decay of super-heavy particles
cannot be a significant part to the observed particle flux, some anisotropy of the arrival
directions of the particles with energies greater than $5.5\cdot10^{19}$ eV has been observed.
The primary objective of the upgrade of the Pierre Auger Observatory will be to answer the
question about the origin of the flux suppression at the highest energies, i.e. the
differentiation between the GZK-effect and the maximum energy of nearby astrophysical
sources. To address all scientific targets, we propose an upgrade of the Pierre Auger Observatory for
improving the physics potential of the data set. 

The aim of the Pierre Auger Observatory is measuring of 
cosmic rays at the highest energies with unprecedented 
statistics and resolution. The first Southern part of the 
Observatory is located in Argentina. It contains
1680 water Cherenkov detector stations distributed over an area 
of 3000 km$^2$ for measuring the charged particles associated 
with extensive air showers (EAS) and 24 telescopes with 
$30 \times 30$ degrees field of view and 12 m$^{2}$ mirror area 
each to observe the fluorescence light produced by the charged 
particles in the  EAS during operation in clear moonless nights. 
The simultaneous observation of EAS by the ground array and the 
fluorescence light called as `hybrid' events improves the 
resolution of the reconstruction considerably and, due to the 
calorimetric nature of the emitted fluorescence light, provides 
energy measurements virtually independent from hadronic interaction 
models.

The surface detector (SD) array of the Pierre Auger Observatory 
started to operate since 2004. The Cherenkov light is detected by
three 9-inch photomultiplier tubes (PMTs) from which the signals of the
anode and last dynode are digitized by 10-bit ADCs. The currently
used Front-End Boards equipped with the ACEX\textsuperscript{\textregistered} \cite{ACEX}
and Cyclone\textsuperscript{\textregistered} \cite{Cyclone}
FPGA are sampled with 40 MHz. However, both FPGA families are already obsolete, 
ACEX\textsuperscript{\textregistered} chips are no more produced.
Data readout of the enhanced surface detector stations will
be facilitated by replacing the current readout electronics by modern state-of-the-art
electronics providing three times faster sampling, a significantly enhanced dynamic
range, and enabling enhanced trigger and monitoring capabilities.

PMTs read out the Cherenkov 
light from the 12 m$^3$ of purified water contained in each
tank. The signals from the anodes (low-gain (LG) channel) and dynodes 
(high-gain (HG) channel) are transported on equal-length shielded cables 
to the Front End Board (FEB), attached as a daughter board to a 
Unified Board (UB). The UB contains a micro-controller that manages 
all processes related to the data acquisition in the detector station.

The splitting of the signals allows an extension of the dynamic 
range of the measured energy range to 15 bits with 5 bits
overlapping. The system digitizes the 6 analog signals of each 
detector station in the multistage differential pipeline 
architecture ADC. 
Signal filtering is performed by anti-aliasing 5-pole Bessel filters. 
The outputs of the six 10-bit ADCs are 
processed by an Altera\textsuperscript{\textregistered} 
FPGA working as trigger/memory circuitry (TMC).
The TMC evaluates the ADC outputs for interesting trigger patterns,
stores data in a buffer memory, and sends an interrupt to the UB if a trigger 
occurs \cite{trigger}.

\section{Requirements for an upgrade}

The electronics, currently used in surface detectors, were designed 10 or more years ago.
10 years in an electronics development is an epoch. Several components are not more produced,
the replacement of the failed components becomes a significant factor.
On the other hand, better and better understanding of fundamental processes imposes
on experiments new challenges requiring higher resolution, faster measurements with
higher accuracy with more sophisticated algorithms etc. 
Fortunately, a significant part of challenges can be accomplished with a new much powerful,
energy-efficient electronics with dedicated, embedded signal processing blocks allowing 
an implementation of much more complicated, mathematical algorithms in real time.

\subsection{Faster timing for the surface detector stations}

The currently used 40 MHz sampling provides rather a smooth digitization.
ADC samples at 25 ns intervals for sure do not extract nano-second details from the analog signal.
However, timing information is crucial for a separation of muon signals from the electromagnetic component.
Very inclined "old" showers with only surviving muon components are very narrow pancake
generating in the water Cherenkov detectors very sharp rising ADC pulses with an exponential attenuation
which can be a signature of the "old" inclined showers. A precision of the rising time measurement is a factor.
On the other hand higher sampling increases a de-synchronization of signals in 3 PMTs decreasing a probability
of 3-fold coincidences used in a standard threshold trigger. A desynchronized signals in a time domain
can be recognized in the Fourier space by i.e Discrete Cosine Transform algorithm
\cite{DCT-NIM}\cite{CycloneIII}\cite{DCT}.

Neutrinos may generate showers starting their development deeply in the atmosphere, known as "young" one.
They contain a significant amount of an electromagnetic component, however, is is usually preceded by a muon
bump. Simulation show \cite{RT2014-ANN} that often it is fully separated from the EM fraction.
However, it this case a very precise timing is fundamental.
Generally, faster electronics provides more detailed ADC profiles sometimes crucial for confirmation or rejection
hypothesis.  The Auger-Beyond-2015 project
assumes 120 MHz sampling. 
A developed new Front-End Board can be operating even with 250 MHz sampling.
In laboratory we detected even 2 ns pulses (see section VII).
However, it should be pointed out that a possibility of 250 MHz sampling in the new FEB is only for tests
to verify (in a possible wide range) several variants of data acquisition. 

A geometrical distance between PMTs is $\sim$1.8 m. Depending on an azimuth angle a differences between an arrival
times of the Cherenkov direct light for horizontal showers reach up to 6-8 ns. With the 25 ns resolution (corresponding
to a currently used 40 MHz sampling) it is rather difficult to get valuable information on the time distribution asymmetries
between PMTs. Nevertheless, for 120 MHz or faster sampling delays between PMTs can be registered as data shifts to the next time bin.
An analysis of relative shifts for ADC profiles in the same surface detector can provide significant hints for a trigger improvement
for very inclined or horizontal showers.

\subsection{Increased dynamic range}

The currently used 15-bit dynamic range with two 10-bit ADCs and 5-bit overlapping is too narrow for
an investigation of showers close to the core, where a saturations even in the low-gain channel appear.
Front-End Board with Cyclone\textsuperscript{\textregistered}  V developed in the University of \L{}\'od\'{z}, 
as a test high-resolution platform for the Auger-Beyond-2015 Front End Electronics,
contains two 14-bit ADCs per PMT
(with 9-bit overlapping giving 19-bit dynamic range) and with two additional channels for potential extra sensors. 
For sure, 14-bit resolution in two HG and LG channels per a single PMT is not necessary in a final design. 
Nevertheless, a large overlapping of HG and LG channels allows a precise calibration between them
for future final design optimization.

\begin{figure}[h]
\begin{centering}
\includegraphics[width=1.0\columnwidth,height=0.6\columnwidth]{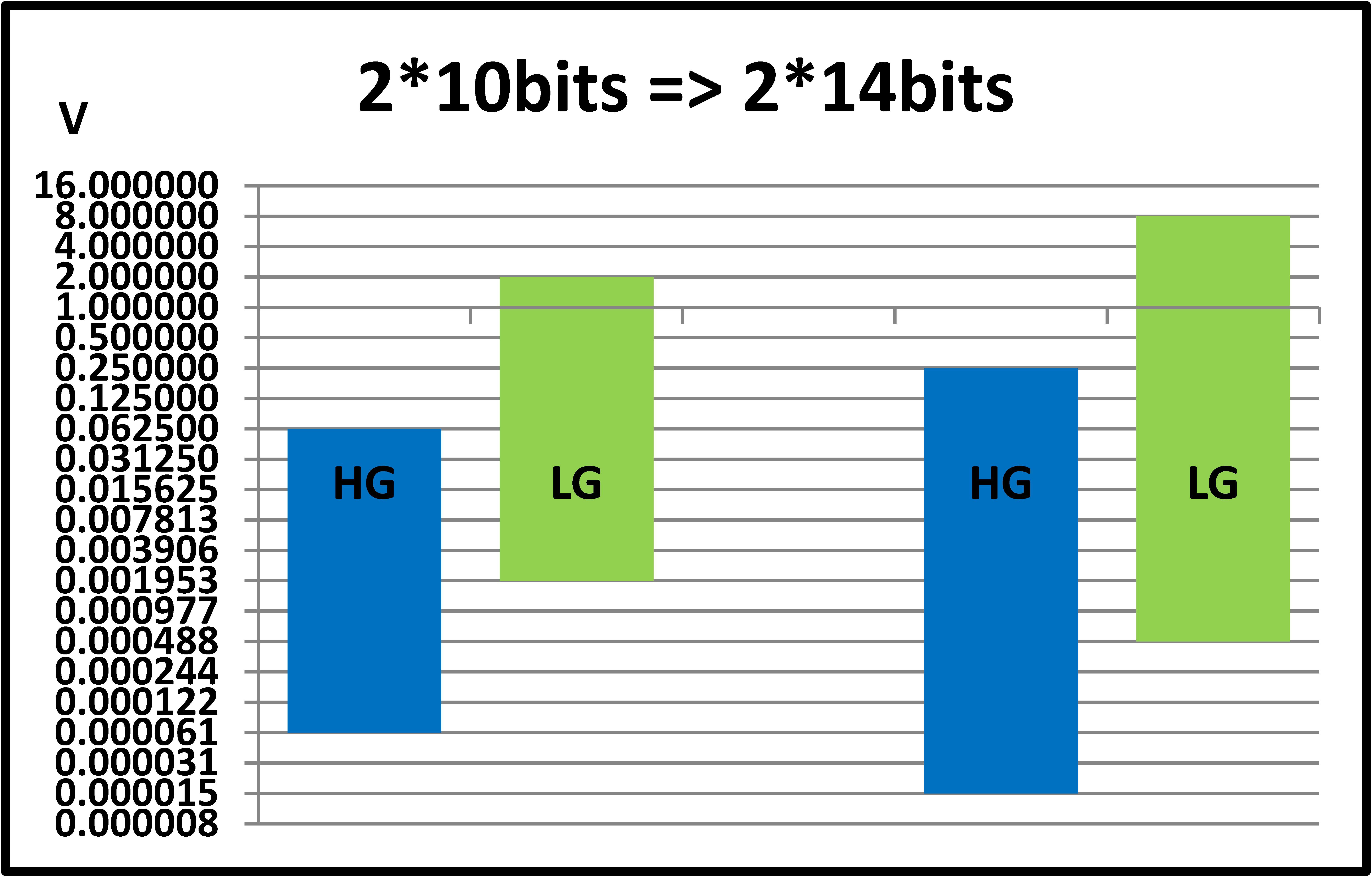}
\caption{Dynamic range in the FEB developed in the University of  \L{}\'od\'{z}, 
with two 14-bit ADCs with 9-bit overlap covering 19-bit dynamic range. 
}
\label{PMT}
\end{centering}
\end{figure}

\section{Analog section}

The ACEX/Cyclone Front-Ends use 6 identical channels driven from PMT dynodes (high-gain channels) or from
PMT anodes (low-gain channels). The amplification in the HG is 32 times higher than in the LG.
In the new design we are going to increase a sensitivity of measurements. Unfortunately, dynodes provide
relatively noisy signals. We will use anodes only and split analog signals directly on the FEB. 
14-bit ADCs in the U\L{} design require a small amplification factor (only x8).
Unfortunately, a total number of the LVDS receivers allows an implementation of 6*14-bit + 2*12-bit ADC.
The proposed electronics upgrade provides a flexible interface to allow the other enhancements 
co-located with the surface detector stations to make use of the
data processing and communications infrastructure of the stations.

In the new design we selected six 14-bit ADCs for 3 HG and 3 LG channels. Actually, 12-bit resolution seems to be enough.
However, taking into account measurements for Auger Engineering Radio Array (AERA) we noticed 
a significant factor of a quantization noise.

We tested an influence of the noise on spectral characteristics short pulses typical for
radio events (sine wave enveloped by a Gauss bell). 
Even for a noise level comparable with a level of signal the signal
spectral contribution was clearly visible and easy to detect. For 3 times higher noise
 we observed a dramatic change of spectra practically making impossible
a signal recognition. This is normal. However for higher signals (with the same signal to noise ratio) 
contributions for very low and very high frequencies are lower in comparison to relatively low signals, 
when quantization of signals additionally affect the spectra. This means that a quantization noise may be reduced
in higher-resolution ADCs due to a tinier granularity \cite{RT2014-wavelet}. 

Nevertheless, this is a prototype which main goal is a test of several sometimes antagonistic conditions.
The final design extracts the best solutions as a compromise between the resolution, speed and the cost.

 \section{FPGA selection}

In a current design Front-End Boards are pluged-into the UB on a 96-pin connector. 
The University of \L{}\'od\'{z} develops an intermediate design of the still plugged-in Front-End Board
with all features required for a final design: a fast sampling, a wider dynamic range, a possibility to merge other
experiments. 

The following criteria were crucial for the FPGA selection:

\begin{enumerate} 
\item{number of LVDS receivers - to receive data from at least 8 ADCs with at least 12-bit (better 14-bit) resolution,}
\item{an amount of equivalent logic elements (LEs) - for an implementation of a general algorithms,}
\item{a capacity of embedded memory - for an implementation of large, fast buffers, typically as dual-port RAM,}
\item{number of variable-precision digital signal processing (DSP) blocks 
that can implement 18x18 embedded multipliers - important for sophisticated algorithms like:
the trigger based on the Discrete Cosine Transform  (DCT) \cite{DCT-NIM}-\cite{DCT} for a detection of very inclined "old" showers,
the trigger based on an Artificial Neural Network (ANN)  \cite{RT2014-ANN} for a detection of "young" showers, 
potentially generated by neutrinos \cite{neutrinos},}
\item{number of fractional clock synthesis phase-locked loops (PLLs) - to create individual PLL clocks 
corresponding to each ADC LVDS clock output,}
\item{a power consumption,}
\end{enumerate} 

All previous generations of the FEB for surface detectors were equipped in Altera FPGAs \cite{ACEX}\cite{Cyclone}.
However, due to a harsh competition between two large FPGA producers: Altera and Xilinx, we considered also
FPGA of the 2nd one, especially the FPGA families integrating the standard fast logic with embedded micro-controllers
(Hard Processing Systems - HPS) and System on Chip (SoC). Both Altera and Xilinx offer chips with HPS and SoC.

Altera's Cyclone\textsuperscript{\textregistered} V FPGAs offer the industry's lowest system cost and power, 
with a high performance levels sufficient for high-volume applications. 
The SoC Altera Cyclone V SE family with ARM\textsuperscript{\textregistered}-based HPS offers the biggest 896-pin FPGA
with only 72 LVDS lines \cite{Cyclone_SE}. Unfortunately, 
it is not enough even for six 12-bit channels (we have to take into account also differential clock lines from ADCs).
Other Altera's family - Arria\textsuperscript{\textregistered} V is too expensive.

The Xilinx Zynq\texttrademark\@-7000 family provides the software programmability of a embedded processor 
with the hardware programmability of an FPGA. 
However, only a single Xilinx SoC FPGA: XC7Z100 offers a large amount  of LVDS lines: 102 for FF900 package or 120 for FF1156 one \cite{Xilinx}.
This chip can support maximal nine 12-bit LVDS ADCs (with FF1156). 4 dual-channel + a single-channel 12-bit ADCs 
require 4*2*12+12=108 LVDS data lines + 5*2=10 lines for differential clocks.

The University of \L{}\'od\'{z} decided to use the FPGA from the Altera's Cyclone\textsuperscript{\textregistered} V E family
without the SoC and HPS as not absolutely necessary for a still plugged-in FEB. The new design focuses on a reliability of the firmware
(LVDS protocol with high sampling, standard and new triggers cooperations, potential lossless data compression, various modes of 
remote and on-side FPGA programming),
optimization of the data acquisition for the final design, measurements of FEB parameters in real Argentinean pampas conditions
as well as on new trigger algorithms.
Additionally, all old firmware written in AHDL for ACEX, Cyclone, Cyclone III and already tested in pampas for years can be easily 
adjusted to the Cyclone V FPGAs.  Xilinx platform, however, does not  support AHDL. The firmware should be written in the VHDL or Verilog
either from the scratch or should be translated from the AHDL. Experiences in a FPGA code development 
suggest to be very careful in simple code translation. Timing in e.g. VHDL is not the same as for the AHDL. 
The AHDL code translated into VHDL indicated a different timing optimization by the Quartus compiler. 
To keep a timing already perfectly verified in real
conditions for years and to save time for additional arduous simulations for timing verification we decided to adjust
the already existing AHDL code for new Cyclone V FPGAs.

Only two chips obey the above criteria : 5CEFA7 (150 kLE, 6.86 Mbit and 312 DSP 18x18 multipliers) 
and 5CEFA9 (301 kLE, 12.2Mbit and 684 DSP 18x18 multipliers). We decided to implement the biggest chip 5CEFA9F31I7
(industrial version with a temperature range -40$^{\circ}$C - +85$^{\circ}$C - due to large daily variation of temperature reaching even 40$^{\circ}$C)
to keep a maximal flexibility in a development of sophisticated, resource consuming algorithms (e.g. ANN for a 
recognition of "young" inclined showers).

\section{FPGA programming}

The surface detector is anticipated to be operating at least 15-20 years. 
In such a long period it is very likely the algorithm of the detection controlling
the FPGA has to be updated. The system should be reconfigurable remotely, if possible.
The currently operating FPGAs in the surface detectors can be reprogrammed via
the Passive Serial mode driven from the external host - the UB.
The prototype FEB based on the Cyclone\textsuperscript{\textregistered}  V E
will be equipped in several programming modes to select the optimal one
in the final version as well as to keep a redundancy in case of some modes failed. 

The prototype FEB is considered as an intermediate version equipped in the 
very sensitive analog section (14-bit ADCs with a standard 120 MHz sampling - 
even 250 MHz sampling is possible) and a powerful
digital trigger/memory circuitry, but plugged into the old UB (a communication 
via DMA at 40 MHz with 1 wait-state). This intermediate FEB will allow a verification
of the new powerful system (both analog and digital sections) without a change
of the existing infrastructure of the data transmission to the Central Data Acquisition Center (CDAS).

The size of the programming files for ACEX\textsuperscript{\textregistered} CPLD
($2^{nd}$ generation - EP1K100Q208I7 \cite{ACEX}), 
Cyclone\textsuperscript{\textregistered} ($3^{rd}$ generation - EP1C12Q240I7 \cite{Cyclone}) or
Cyclone\textsuperscript{\textregistered} III ($4^{th}$ generation - EP4C40F324I7 \cite{CycloneIII}) 
were on the level of 2*180kB, 150kB and 440kB, respectively. However, the size of the rbf file
for the largest chip from the Cyclone\textsuperscript{\textregistered} V family - 5CEFA9F31I7N
(selected for the design)  reaches 12MB. 

The Cyclone\textsuperscript{\textregistered} V E could be programmed from the UB level (as it 
receives  the configuration file from the CDAS via radio). However, the UB contains only 2MB of RAM
for a full operational system. The programming file has to be transmitted in pieces and stored
in a nonvolatile memory. The transfer speed via a radio is only 2400 bps. The 12MB file will be transmitted 
more than 14 hours. It is too long time to be wasted only for a new file transmission. The 
transmission is considered in a background without a break of a normal operation. The used 
nonvolatile memory has a capacity 128MB, large enough to store 8 different configuration files.
The prototype is equipped in several programming modes in order to increase a reliability of the entire system.
All previous generation of the FEBs used only a single mode - Passive Serial, the configuration file
was uploaded to the FPGA from the external host (UB). The FPGA in the current design can be programmed
in a following modes:

\begin{enumerate} 
\item {the Passive Serial via 10-bit connector and the USB-Blaster programmer for laboratory tests - required 
reprogramming after power down,}
\item{the JTAG, considered for the laboratory tests for a temporary programming of the FPGA,}
\item{the JTAG + Serial Flash Loader, considered for the laboratory tests, which also allows to program 
the configuration device EPCQ256,}
\item{combining JTAG programming of configuration device and FPGA with Active Serial configuration of FPGA using 
the EPCQ256 and the USB-Blaster (the MSEL[4..0] pins cannot be driven from the digital sources, they have to be
connected directly to the GND or VCC, a dynamical switch between Active and Passive Serial modes is not allowed),}
\item {the Passive Serial (as in the previous generations of the FEBs) via the MAX V CPLD, 
the configuration file is uploaded sequentially from the UB to the serial NOR Flash memory N25Q00AA13GSF40F
and next reloaded in a single shot from the memory to the FPGA.}
\item{the Passive Serial via MAX V CPLD using the SDHC card (2 GB) located on the daughter board plugged into the
10-pin connector. This mode is considered as the main one for the field operation. During the SD maintenance
on the field, the local staff has $\sim$30 min for a single SD, it is too short to upload several config files via standard UB serial 
port even with 115.2 kbps speed. UARTs in the NIOS also support max. 115.2 kbps only. 
The SDHC card will be programmed in the laboratory and only plugged into the
connector on the pampas. A single command from the CDAS is enough to reprogram the FPGA: the command selecting
the address of the config file in the SDHC card.}
\end{enumerate}

\begin{figure}[t]
\begin{centering}
\includegraphics[width=1.0\columnwidth,height=0.7\columnwidth]{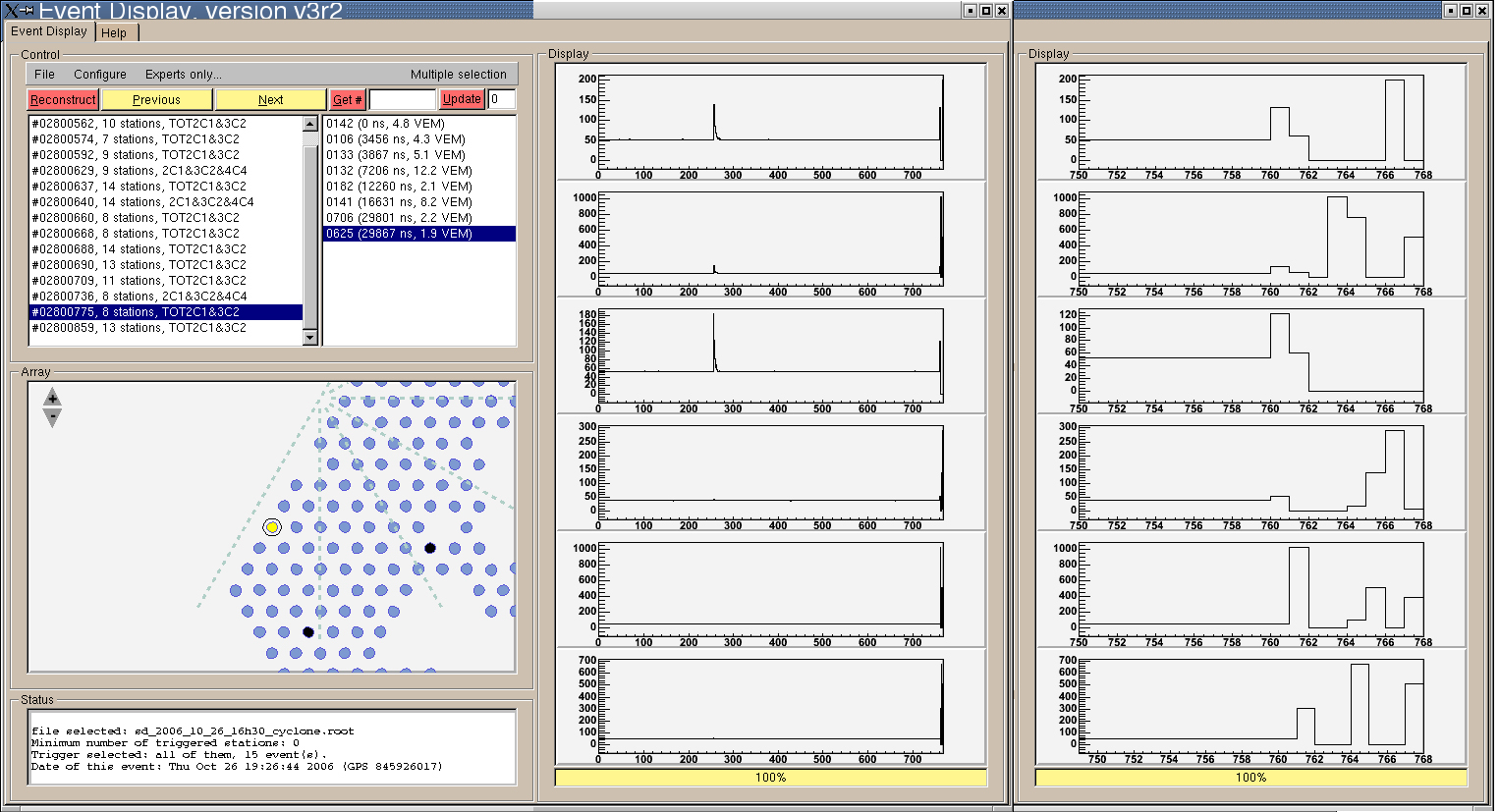}
\caption{ADC traces with inserted diagnostic data on the last 8 words of the trace, where always noise appears .}
\label{diagnostic}
\end{centering}
\end{figure}

\section{FPGA code}

The FPGA code has been developed by the author for the APEX \cite{APEX}, ACEX \cite{ACEX}, 
Cyclone \cite{Cyclone} and Cyclone III \cite{CycloneIII} designs.
Data from the FPGA are transferred via DMA protocol to the UB and next analyzed by the T2 software trigger
via radio to the CDAS. Data are available for users in root files. Each trace for the events contains data 256
time bins before the trigger and 512 time bins after the trigger. For 40 MHz sampling frequency it takes a profile
of the signals for 19.2 $\mu$s. Analysis of huge amount of traces showed that last 8 words in each trace contain
practically only noise. Monitoring files collect informations on the working system, however, they cannot register
specific information related to the particular events. These informations have been inserted on the end of the
real traces in so called "diagnostic" mode.  For one year (2010) 6 surface detectors were working with the 
diagnostic mode (Fig. \ref{diagnostic}) on the Engineering Array.

\begin{figure*}[t]
\begin{centering}
\includegraphics[width=2.0\columnwidth,height=1.3\columnwidth]{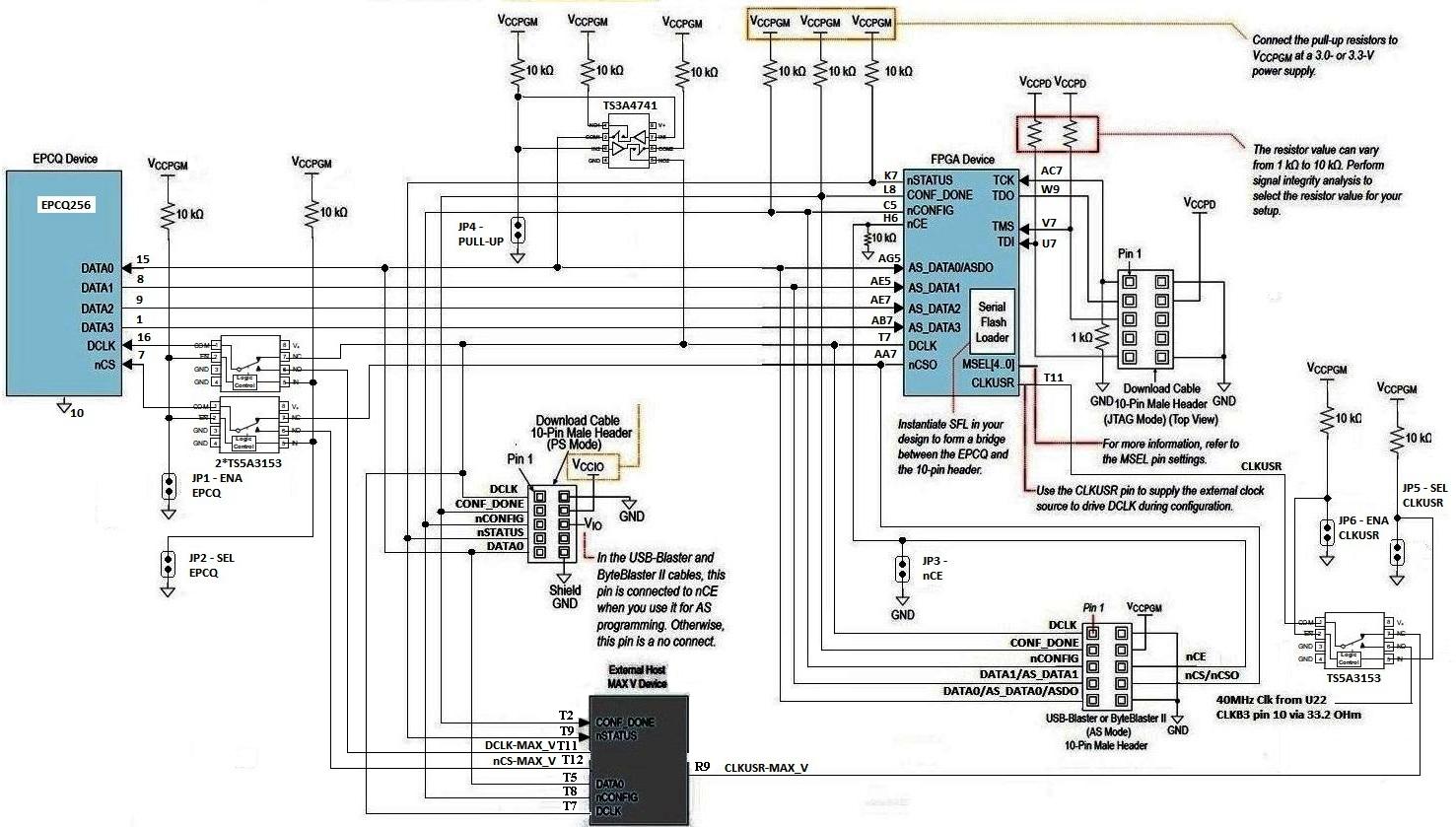}
\caption{Schematics of multi-config circuit in Passive and Active Serial as well as JTAG modes.
Nonvolatile EPCQ memory supports In System Programming mode.}
\label{config}
\end{centering}
\end{figure*}

The currently used FEB provides 6*10 bits data for 768 time bins for each event. 60*768 matrix is a standard
format for SD data. Higher resolution, higher sampling, longer traces require a new format, however, 
it is a task for the UUB which is currently in a development phase. This intermediate FEB has to use the standard
data format in order not to change any cell in a data transmission chain.

The trigger in the surface detector array is hierarchical. Two levels of trigger (T1 and T2) are used.
T2 triggers are combined with those from other detectors and examined for spatial and temporal correlations, 
leading to an array trigger -T3 which initiates data acquisition to the CDAS.
Data for each triggered event is stored on the local UB memory for 10 s waiting for a possible T3.
Two independent trigger modes are implemented as the T1, having been conceived to detect, 
in a complementary way, the electromagnetic and muonic components of EAS. The
first T1 mode is a simple threshold trigger (Thr) which requires 3-fold
coincidence above 1.75 $I_{peak}^{VEM}$. This trigger is used to select large signals that are not spread in
time. Thr trigger is particularly effective for the detection of very inclined showers
which form a flat pancake of muons and generate sharp narrow ADC traces.
This trigger reduces the rate due to atmospheric muons from  3 kHz to 100 Hz. 
Thr trigger rate is suppressed in the T2 trigger to $\sim$20 Hz. The 2nd T1 mode - 
ToT (Time over Threshold) is used for a detection of particles and
photons at the detector dispersed in time. ToT rate is on the level of 2 Hz and is not suppressed by the T2.

\begin{figure}[t]
\begin{centering}
\includegraphics[width=1.0\columnwidth,height=0.75\columnwidth]{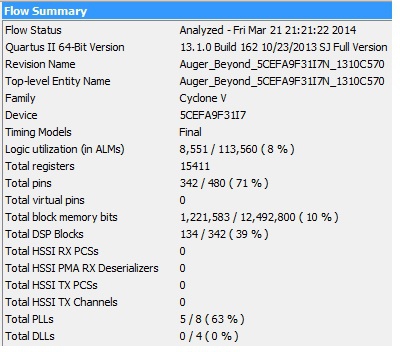}
\caption{Compilation report from theQuartus II v.13.1 compiler for 160 MHz sampling code with 
the standard Thr and ToT triggers as well as the DCT trigger driven by 3 engines for each PMT.}
\label{quartus}
\end{centering}
\end{figure}

\begin{figure*}[t]
\begin{centering}
\includegraphics[width=2.0\columnwidth,height=1.2\columnwidth]{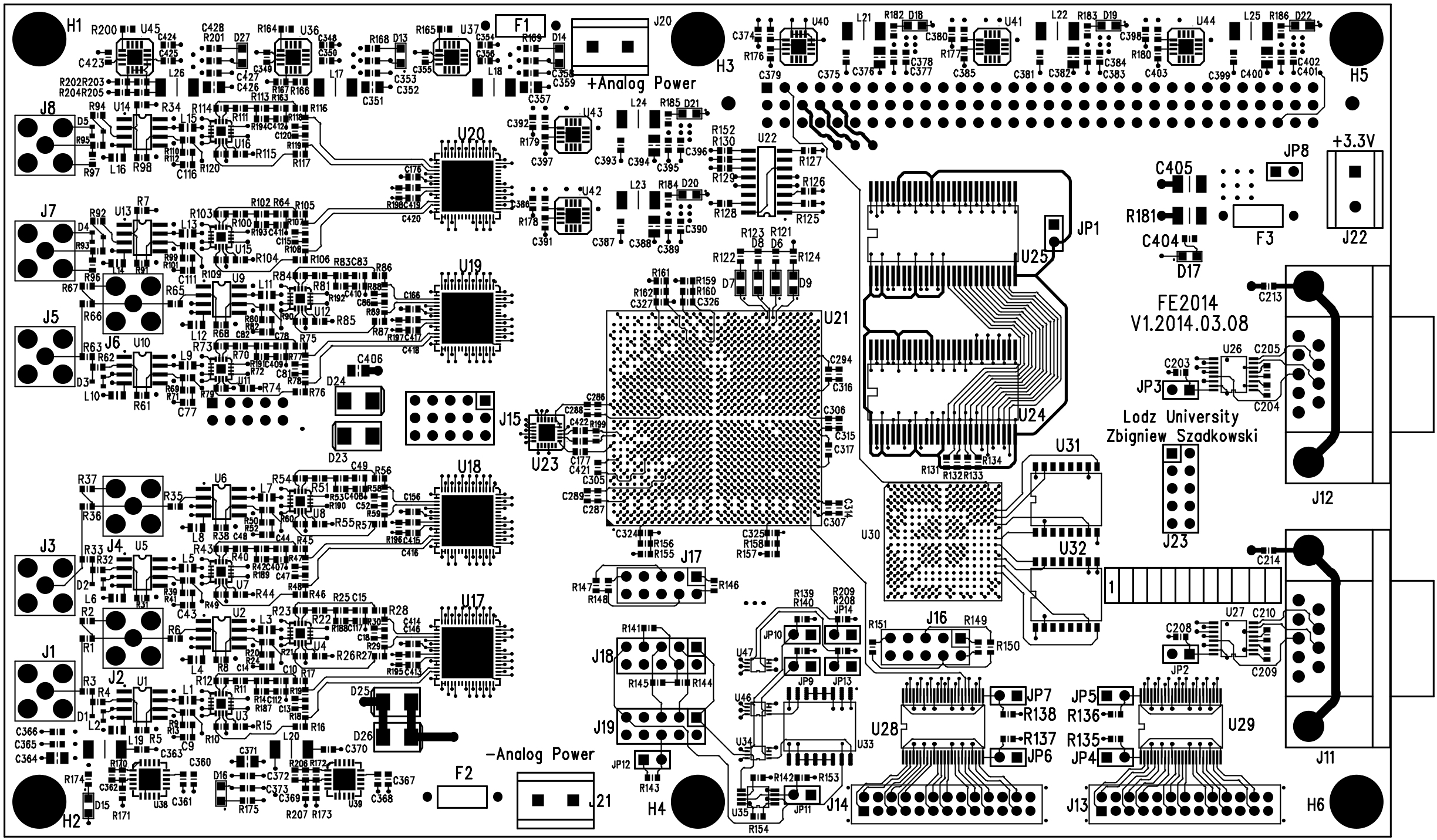}
\caption{The top layer of the fabricated Front-End Board (12 layers). 
Due to a very long delivery time of some components the photo of the board will be available in Oct. 2014.}
\label{top}
\end{centering}
\end{figure*}

\begin{figure}[t]
\begin{centering}
\includegraphics[width=1.0\columnwidth,height=0.6\columnwidth]{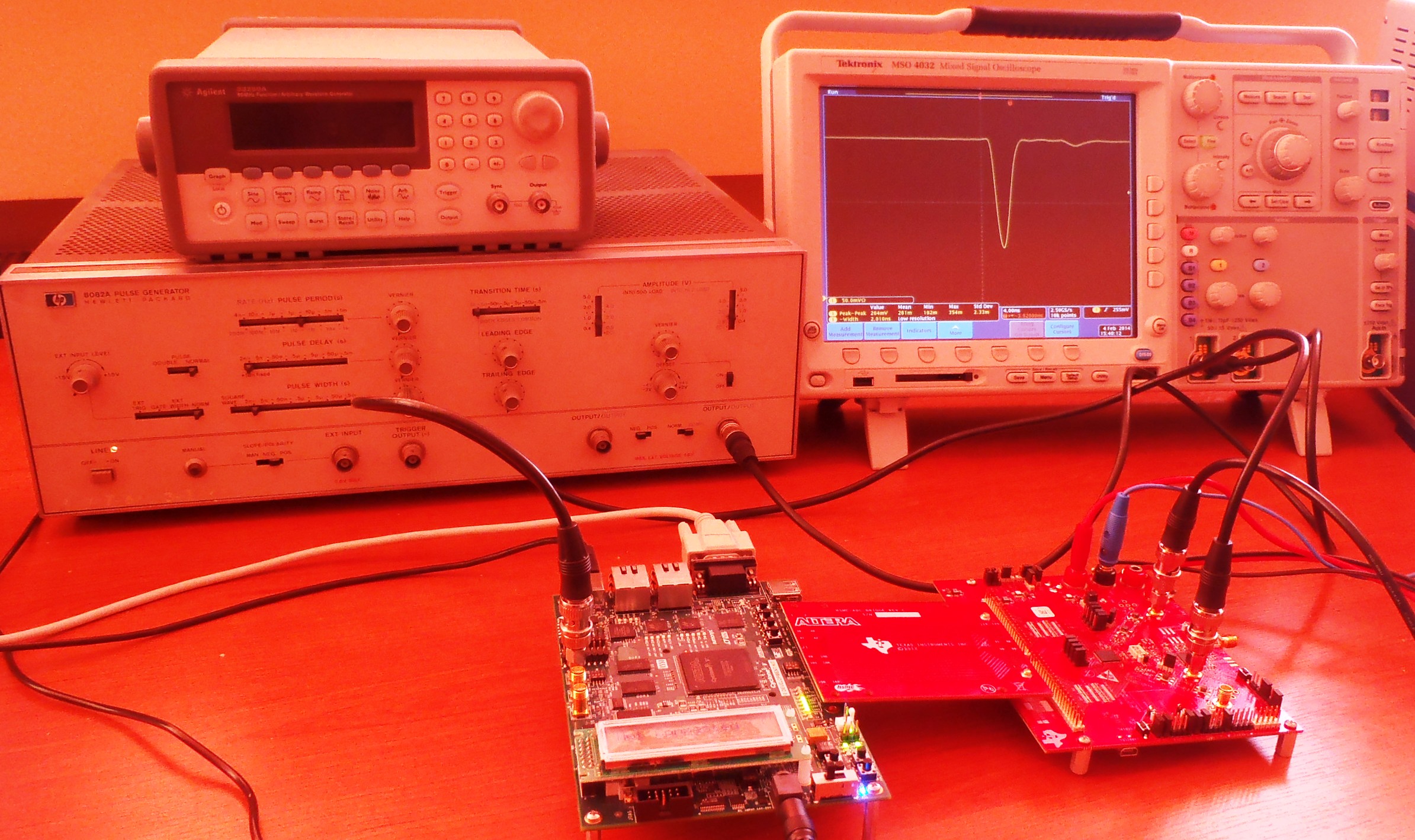}
\caption{The laboratory setup with the Altera development kit DK-DEV-5CEA7N (Cyclone V E FPGA)
Altera HSMC-ADC-BRIDGE and Texas Inst. ADS4249EVM Evaluation Module. Here, the HP8082A generates
very short 2 ns pulse - visible on the Tektronix MSO4032 scope.}
\label{setup}
\end{centering}
\end{figure}

6 channels for 14-bit ADC provide 84 bit data for each time bin. 160 MHz sampling reduces the time window
from 19.2 $\mu$s to 4.8 $\mu$s only for 768 time bins. Only HG channels participate in a trigger generation.
Several configurations of data format (compatible with the standard one - not to violate the interpretation 
of the event by the T2, T3 and the CDAS) will be tested. All additional informations will be inserted into 8 last words
of the trace (the diagnostic mode). 10 low significant bits from the 14-bit HG ADC are put on the place of standard 10 bits
of the HG channel. 
If 14-bit value is grater than 1023 (10 LSB may be even zeros) these 10 LSB are artificially set to \#3FF 
to inform the T2 on the saturation in the HG channel. 
The LG channels can contain e.g. either:
\begin{itemize}
\item{4 high significant bits of the HG ADC put on the 4 low significant bits of the LG channel + 6 most significant 
(but non zeros) bits from LG 14-bit ADC or}
\item{10 bits from the LG 14-bits ADC with adjusted position to neglect zeros for high significant bits if signal is small}
\end{itemize}

Such a format provides non-overlapped information for strong signals as well as a possible dynamically overlapped 
information for relatively small signals.

4.8 $\mu$s window can be too narrow especially for events registered far from the shower core which are spread on time.
The FPGA stores event data in two switching buffers (to reduce a dead time) with 4 times longer length (1024 time bin
before a trigger + 2048 time bin after it). The length of the buffer corresponds to 19.2 $\mu$s. The additional FPGA
procedure checks a signal contribution for [0..768] and [1536..3072] time bins (it means earlier than 256 time bins before
the trigger and later than 512 time bins after the trigger). If a significant signal contribution were found, data can be transmitted
either:
\begin{itemize}
\item{as sum of 2 or 4 neighboring time bins or }
\item{with additional 1 or 3 transmissions of "extended" event. }
\end{itemize}

\begin{table}
\centering\caption{\label{Voltage} The list of needed analog inputs and digital I/O lines.}
\begin{tabular}{|c|c|c|c|c|}
\hline
   Bank  &  Total  &  Used & Voltage& Function \\
               &   Pins  &  Pins  &                   &                   \\
\hline \hline                                                   
   8A       & 80  & 40 &  2.5 V  &  40 LVDS inputs for channels 1 - 3\\ \hline
   7A       & 80  & 40 &  2.5 V  &  40 LVDS inputs  for channels 3 - 6  \\   \hline
  6A        & 80  & 40 &  2.5 V  &  40 LVDS inputs  for channels 6 - 8               \\         \hline
  5B        & 48  &  36 & 1.8 V  &   lines controlling the ADCs at 1.8 V          \\     \hline
  5A        & 32  &  32 & 3.3 V &    I/O data lines between FPGA and UB     \\    \hline
  4A        & 80  &  78 & 3.3 V &   Ctrl lines (FPGA $\Leftrightarrow$ UB) + SDRAM        \\         \hline
  3B        & 48  &  45 & 3.3 V &   Cyclone V  $\Leftrightarrow$ MAX V + 2*UART\\        \hline   
  3A        & 32  &  31 & 3.3 V &   Cyclone V  $\Leftrightarrow$ MAX V\\ \hline
\end{tabular}
\end{table}

The 1st variant provides a loss compression, the 2nd one offers lossless data transfer, however, it requires
a significant software modification in the UB and the CDAS. The 2nd variant is treated as the possible backup.
However, such a format modification has to be considered because data from two 12-bit ADCs (i. e. from the SPMT) 
should be also transmitted in a final design.

The FPGA code receives data on totally 108 (6*14+2*12) LVDS lines. Cyclone V families support OCT (On Chip Termination)
option. No any external 100 $\Omega$ resistors for LVDS termination is needed. The ADS4249 ADCs provides the outgoing clock
driving the PLL in the FPGA for the optimal LVDS data receiving. 4 PLL circuits (each for double ADCs in ADS4249) with individual 
PLL clocks drive 3 synchronization register stages before the global FPGA clock. 

The Cyclone V FPGA implements also the NIOS processor supporting an external SDRAM (128MB), two RS232 UARTs and two 16-bit I/O buffers
for additional experiments (see section III). 
Fig. \ref{quartus} shows the compilation report for 160 MHz sampling and all 8 ADC channels with LVDS receivers, 3 DCT engines
and the NIOS processor. DCT engines \cite{DCT} contain sigma-delta algorithm for even huge daily temperature variation
\cite{atmospheric}.
Each bank of the Cyclone V E can be supplied and drive I/O with various voltage standard. 
Table \ref{Voltage} lists I/O lines connected to the Cyclone V E banks. 

\begin{figure}[h]
\begin{centering}
\includegraphics[width=1.0\columnwidth,height=1.0\columnwidth]{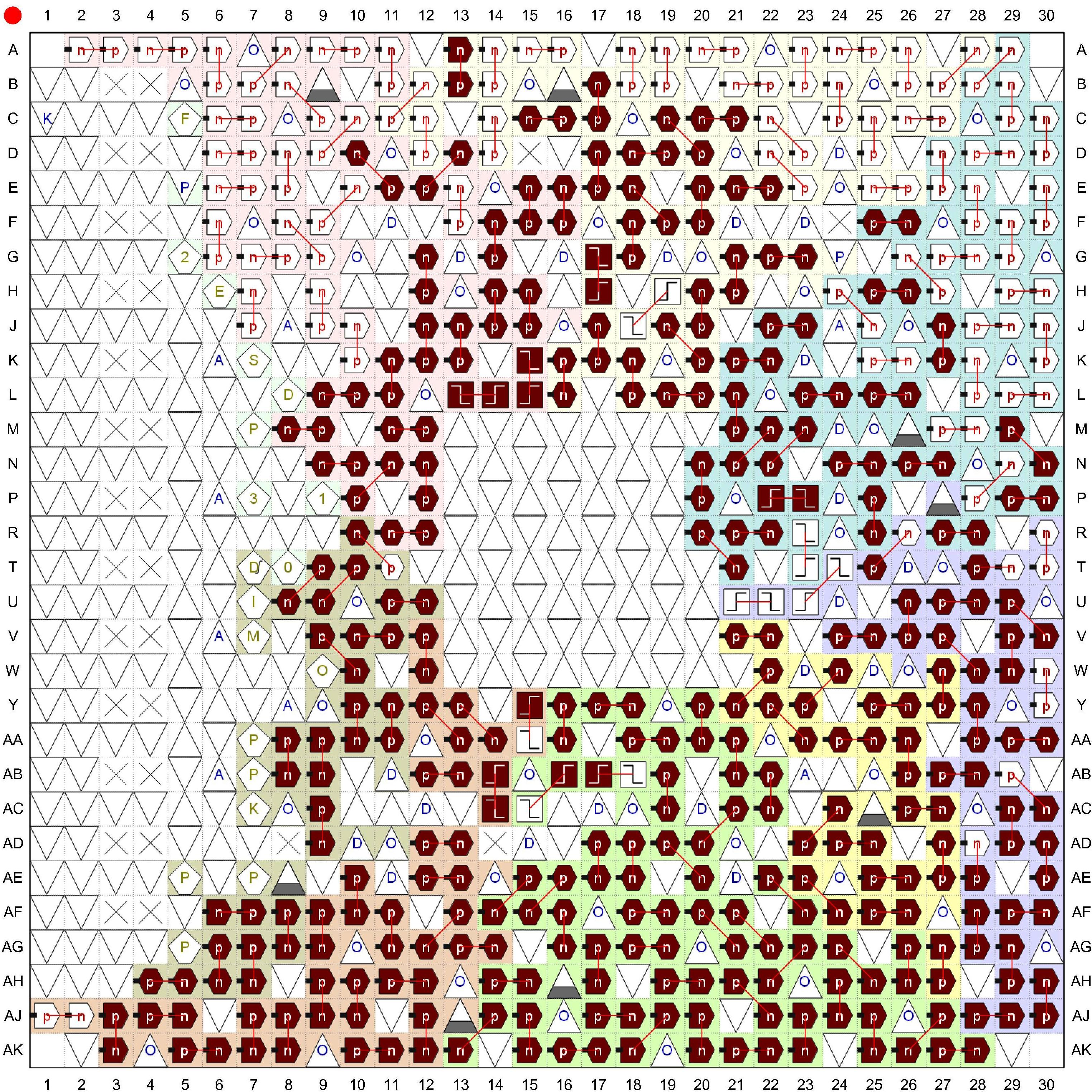}
\caption{Pins assignment for BGA896 package of Cyclone V FPGA 5CEFA9F31I7. All LVDS receivers are used. 
Free unassigned pins correspond mainly to ont used LVDS transmitters.}
\label{pin}
\end{centering}
\end{figure}

The previous version of the Quartus II v. 13.0 did not report any warnings for the FPGA code. The latest version, however,
Quartus v. 13.1 reported for the previous code hundreds critical warnings due to cross-talks between LVDS and single-ended lines. 
The reasons was a relatively small distances between
LVDS pin inputs and single-ended CMOS outputs. Movements of pin assignments with a gap of 2-3 pins removed all warnings.
Nevertheless, it is a hint to select a FGPA chip with large amount of pins (e.g. BGA896 package instead of F672, U484 or U324 cheaper
versions) to provide a sufficient flexibility in pins assignments (Fig. \ref{pin}).

\section{Design optimization in laboratory tests}

The proposed sampling frequency for the new FEB is 120 MHz. However, for a prototype we developed
much faster both analog electronics with a digitizer and the FPGA code. For a development we used the Altera
development kit with Cyclone V E FPGA 5CEFA7F31I7. As the ADC we used Texas Instr. ADS4249EVM (Evaluation Module)
with double channel 14-bit ADC with max. 250 MHz sampling. ADS4249EVM send data to the 5CEFA7F31I7 via
Altera HSMC-ADC-BRIDGE in the LVDS standard. Connection between the ADC and the FPGA was a length
of $\sim$20 cm and operated with very high reliability. Outgoing ADC clock supported next the internal PLL in the FPGA.
Each PLL for a single ADC chip (each 2 channels) optimized reading LVDS interleaving data in a middle of a stable region.  
Even for 20 cm distance the LVDS transmission was perfect. In a developed FEB a distance between the ADCs and the FPGA 
is much shorter (Fig. \ref{top}). Input signals were tested with 200 MHz sampling.

\begin{figure}[t]
\begin{centering}
\includegraphics[width=1.0\columnwidth,height=0.6\columnwidth]{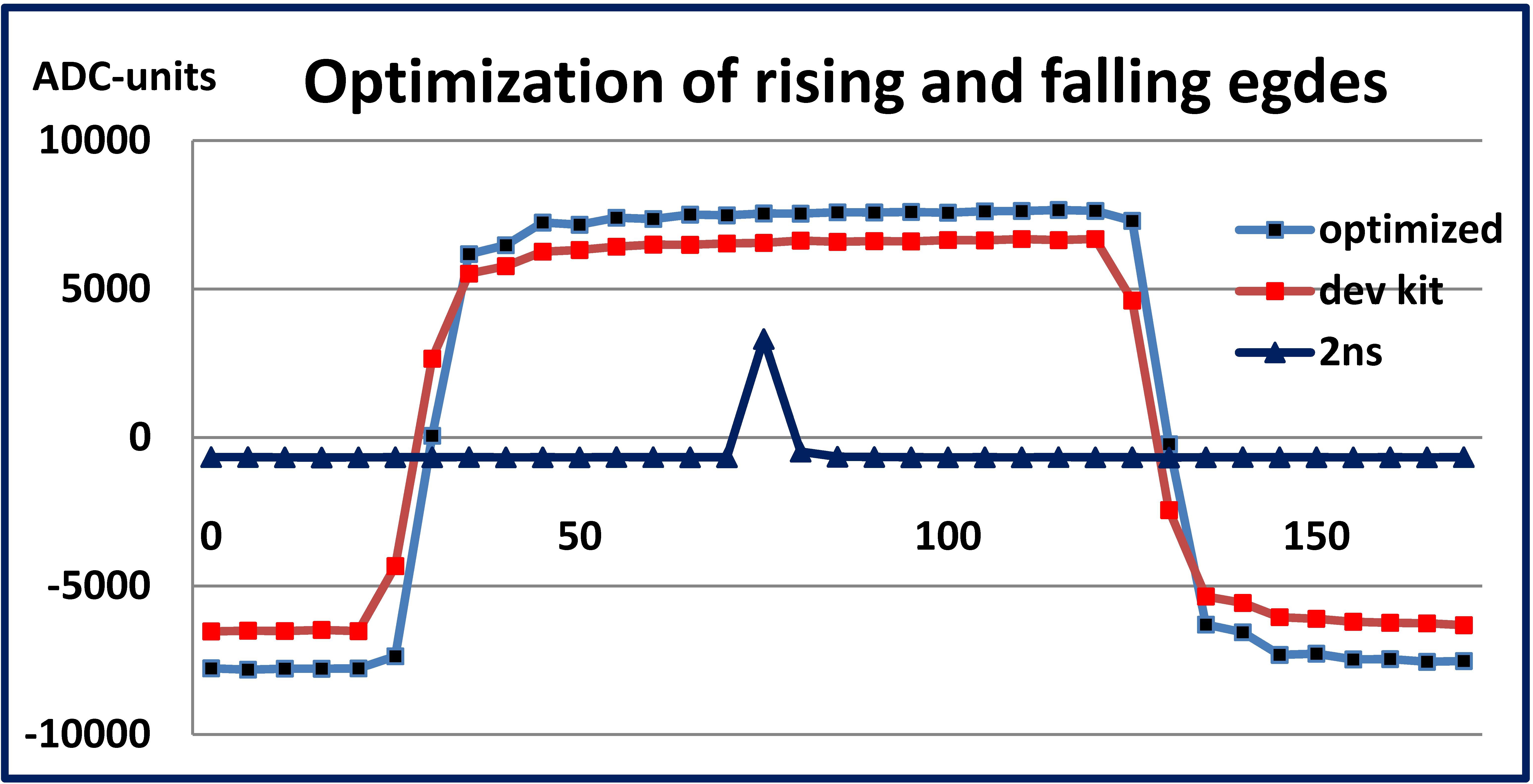}
\caption{An example of registered pulses before and after optimization of the differential driver.
After optimization edges are sharper. The system can recognize even 2 ns pulses. }
\label{pulses}
\end{centering}
\end{figure}

The analog section on the ADS4249EVM contains high-frequency transformer and THS4509 differential drivers.
Transformers differentiate signals and cannot be used. The differential driver does not support DC-DC connections.
All serial capacitors have to be removed. DC-DC connection, however, required a significant tuning of RC components to
get  sharp enough rising and falling edges. Fig. \ref{pulses} shows sample pulses before and after optimization. 
The 2 ns very short pulse (showing on the scope (Fig. \ref{setup}) is read in a single time bin above a pedestal (Fig. \ref{pulses}).

The RMS measured for long traces is on a level of 2.45 ADC-unit (14-bit data processing).
Nevertheless, let us notice that the noise was measured for non-standard configuration:
The ADC connected to the FPGA via ADC-HSMC-BRIDGE with two HSMC connectors
(total distance $\sim$20 cm) with manually assembled resistors and capacitors for high-frequency
response optimization. We expect much better final characteristics in the final FEB.
The test setup, developed originally for an optimization of the upgrade design, has been used also for
test of the FIR filter based on the linear predictor to suppress the RFI contaminations in the radio
detectors of the AERA \cite{AERA} \cite{VCI} \cite{RT2014-LP}.

\begin{figure}[t]
\begin{centering}
\includegraphics[width=1.0\columnwidth,height=0.7\columnwidth]{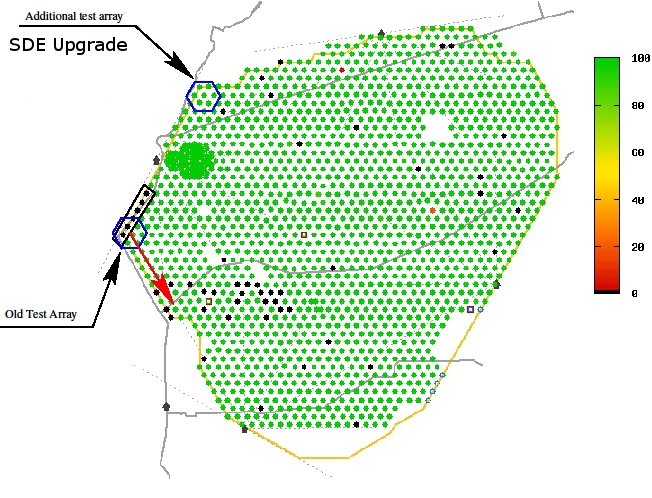}
\caption{Location of the SDE Upgrade Engineering Array on the Pierre Auger Surface Detector Array.}
\label{Array}
\end{centering}
\end{figure}

\section{Conclusion}

 The Technical Board of the Pierre Auger Collaboration
selected 8 surface detectors (hexagon + twin in the center for an investigation of possible GPS jitter)
in a nord-west region of the SD array (Fig. \ref{Array}) for tests of the new FEB on the Cyclone V platform 
with 3-4 time higher sampling and 14-bit resolution with a cooperation of the SPMT and possible other detectors. 
Simultaneously, the DCT triggers will be implemented parallel with the standard ones to verify the detection of very inclined showers
based on an online analysis of a shape of signals in a frequency domain.

10 FEBs were ordered for a preproduction. Unfortunately, a delivery time for Cyclone V FPGA and some other components
reaches 80 working days. Deployment of prototype FEBs and tests in real pampas conditions are expected in the 2nd part of 2014.

We plan tests for several months to verify 14-bit and 12-bit DAQ (2 LSBs turned-off for future cheaper 12-bit ADCs),
various variants of data transmission to the CDAS in the current narrow radio channel (dynamical selection of significant range of data
for lossless DAQ - both in an amplitude and a time domain or dynamical summation of neighboring time-bins for large signals,
when details cease being significant), new trigger algorithms and e.g. NIOS support for temporary data storage.

We believe that data obtained from the intermediate FEBs provides a significant improvement of the final design for the
Auger-Beyond-2015 which really allows new discoveries in astrophysical researches.

\section*{Acknowledgment}

The author would like to thank Yury Kolotaev from the Siegen University for
the PCB design.

This work was supported by the Polish National Center for
Research and Development under NCBiR Grant No. ERA/NET/ASPERA/02/11
and by the National Science Centre (Poland) under NCN Grant No. 2013/08/M/ST9/00322

\ifCLASSOPTIONcaptionsoff
  \newpage
\fi

\end{document}